\documentclass[aps,prl,twocolumn,showpacs,superscriptaddress,groupedaddress]{revtex4}  % for review and submission

\usepackage{fancyhdr}
\usepackage{amsmath,amsthm,amssymb}
\usepackage{enumerate}
\usepackage{color}
\usepackage{graphicx}
\usepackage{caption}
\def\L{\mathcal{L}}

\def\beq{\begin{equation}}
\def\eeq{\end{equation}}
\def\bea{\begin{eqnarray}}
\def\eea{\end{eqnarray}}

\def\bwt{\begin{widetext}}
\def\ewt{\end{widetext}}

\begin{document}
\author{Mehdi Saravani}
\affiliation{School of Mathematical Sciences, University of Nottingham, University Park, Nottingham, NG7 2RD, UK}

\title{Dark matter and nonlocality of spacetime}
%\date{}

\begin{abstract}
In this letter, we review a candidate for dark matter, known as O$f$DM, and explain how it relates to spacetime nonlocality. This connection provides a physical interpretation for why O$f$DM exists. Given the state of direct and indirect searches for dark matter, O$f$DM model would be an important candidate to consider, since it predicts all direct searches would fail to detect DM. The model has only one free parameter and in this regard, it is highly predictive. We review a few avenues to test this model.  
\end{abstract}

\maketitle

Without any doubt, dark matter (DM) is one of the most important problems in modern physics. After many years of discovering its first evidences, our knowledge of what constitutes DM is very limited and restricted to its gravitational interactions. That is partially the reason for a wide range of candidates for DM, from modifying the gravitational theory to weakly interacting massive particles, axions, etc. 
Despite many direct and indirect searches though, DM has not been detected yet. These experiments have already ruled out many interesting DM candidates or pushed them to the limits.

In this letter, we review a new DM candidate, named O$f$DM (off-shell dark matter) in line with earlier terminology \cite{Saravani:2016enc}, that is consistent with all searches for dark matter. To this end, O$f$DM will not be detected in any direct or indirect search for DM.

What makes O$f$DM an interesting candidate is the fact that it explains why DM exists. In other words, it explains why a form of matter exists that is only visible to us via its gravity, and it connects DM to a fundamental concept in physics, namely the spacetime nonlocality. Originally, the idea of O$f$DM has been developed by studying nonlocal field theories arising from continuum limit of wave propagations on causal sets \cite{Saravani:2015rva}. Causal Set \cite{Bombelli:1987aa} is an approach to quantum gravity which replaces the continuum spacetime with a discrete Lorentzian structure. 

There is a more concise description of O$f$DM in terms on nonlocal fields. We postpone introducing this formulation at a later section. For now, we describe the core idea behind O$f$DM using an equivalent local field theory formulation, without referring to the spacetime nonlocality, and provide a prescription on how to introduce O$f$DM in a theory. In order to introduce this prescription, we will make certain choices and assumptions which more or less looks {\it arbitrary} at this point. The reasoning behind making these assumptions becomes clear in the later sections.

In this letter, we use the term ``matter'' to refer to particles and fields that are visible to us, as opposed to dark matter.
%%%%%%%%%%%%%%%%%%%%%%%%%%%%%%%%%%%%%%%%%%%%%%%%
%%%%%%%%%%%%%%%%%%%%%%%%%%%%%%%%%%%%%%%%%%%%%%%%
\section{Introduction to O$f$DM}
%%%%%%%%%%%%%%%%%%%%%%%%%%%%%%%%%%%%%%%%%%%%%%%%
%%%%%%%%%%%%%%%%%%%%%%%%%%%%%%%%%%%%%%%%%%%%%%%%
In this section, we explain a prescription to introduce O$f$DM in a theory. Moreover, we argue why O$f$DM is a viable DM candidate.

Let us consider a very simplified version of the universe that matter is made of a single massless scalar field $\phi_0$
\beq
\L_M = \frac{1}{2}\phi_0 \Box \phi_0 - V(\phi_0),
\eeq
where $\L_M$ is the matter Lagrangian and $V(\phi_0)$ is any interaction involving $\phi_0$.

In order to include O$f$DM, we introduce a series of infinitely many scalar fields $\phi_k,~ k=1,2,\cdots$with masses $m_k^2 = k \Delta$ where $\Delta$ is a small separation scale in the square of mass and the theory is defined in $\Delta \rightarrow 0$ limit. $\phi_k$ fields play the role of DM as will be explained. Moreover, any interaction involving matter ($\phi_0$) is mediated through an auxiliary field $\phi$ where
\beq\label{phi_def}
\phi = \phi_0 + \sum_{k=1}^{\infty}\sqrt{\tilde\rho_k \Delta}\phi_k,
\eeq
and $\rho_k$'s are positive numbers. The physical interpretation of $\tilde\rho_k$ and why the combination $ \sqrt{\tilde\rho_k \Delta}$ appears in eq. \eqref{phi_def} will be clear in the next section where we explain the nonlocal description of the theory. For now, following our prescription we arrive at the complete Lagrangian of matter and DM as
\beq\label{Lagrangian}
\L =  \frac{1}{2}\phi_0 \Box \phi_0 + \sum_{k=1}^{\infty} \frac{1}{2}\phi_k(\Box-m_k^2)\phi_k - V(\phi),
\eeq 
where $\phi$ is defined in eq. \eqref{phi_def}. This Lagrangian includes O$f$DM fields ($\phi_k,~ k=1,2,\cdots$) and the interaction between matter ($\phi_0$) and DM, $V(\phi)$. Unless explicitly mentioned, $k=1,2,\cdots$ and excludes 0. Now let us explain why $\phi_k$'s are viable DM candidate.

In order to illustrate this point, let us consider a simple interaction like $V(\phi)=\lambda \phi^4$ and the scattering of two $\phi_0$ particles at the tree level. The outgoing particles can be $\phi_0\phi_0$, $\phi_0\phi_k$ and $\phi_k\phi_{k'}$. Consider each scattering cross section ($\sigma$) separately and how they depend on $\Delta$. 

The total cross section of $\phi_0\phi_0\rightarrow\phi_0\phi_0$ is independent of $\Delta$, while 
\bea
\sigma(\phi_0\phi_0\rightarrow\phi_0\phi_k) &&\propto \tilde\rho_k\Delta,\\
\sigma(\phi_0\phi_0\rightarrow\phi_k\phi_{k'}) &&\propto \tilde\rho_k\rho_{k'}\Delta^2.
\eea
For a fixed $k$ (and $k'$), the above cross sections vanish as $\Delta\rightarrow 0$. However, if we sum over $k$ (and $k'$) to obtain the total cross sections we would get a non-zero contribution. In fact, we would recover an integral over mass of the form $\int \cdots \tilde\rho(m^2)dm^2$ where $\tilde\rho(m_k^2)=\tilde \rho_k$ and $\cdots$ corresponds to proportionality factors in the equations above. 

The physical interpretation of this result is straightforward. The interaction with an individual $\phi_k$ field is vanishing as $\Delta \rightarrow 0$. However, at the same time the number of all $\phi_k$ fields is growing. These two factors balance each other to result into a total non-vanishing finite cross section.  This means, although the cross section to produce individual $\phi_k$ particles is infinitesimal, the total cross section to produce all $\phi_k$'s (DM) is non-zero.
In other words, as $\Delta\rightarrow 0$ any interaction involving $\phi_k$'s would be infinitesimal but this is compensated by the increase in the number of $\phi_k$ fields. This is the mechanism by which we can produce DM from matter particles in our universe.

So far we have considered scatterings to produce O$f$DM particles. Now, let us consider scatterings involving a $\phi_{k^*}$ incoming particle. One can verify that, e.g.
\bea
\sigma(\phi_{k^*}\phi_0\rightarrow\phi_0\phi_0) &&\propto \tilde\rho_{k^*}\Delta,\label{cross_section_1}\\
\sigma(\phi_{k^*}\phi_0\rightarrow\phi_k\phi_0) &&\propto \tilde\rho_{k^*}\tilde\rho_{k}\Delta^2,\\
\sigma(\phi_{k^*}\phi_0\rightarrow\phi_k\phi_{k'}) &&\propto \tilde\rho_{k^*}\tilde\rho_{k}\tilde\rho_{k'}\Delta^3.
\eea
The total cross section of all these processes would vanish as $\Delta\rightarrow 0$ even after summing over $k$ (and $k'$), because there is always one factor of $\Delta$ (associated to $\tilde\rho_{k^*}$) that makes the total cross section vanish. This comes from the fact that there is no summation over $k^*$, as this corresponds to an incoming particle. The above argument holds as long as there is at least one $\phi_k$ particle in the incoming state and it explains why DM particles are not detectable directly in scattering experiments. Moreover, it shows that O$f$DM particles are stable and do not decay. Note that the above argument is valid for any type of interaction and is not limited to $\lambda \phi^4$. The key part holding this argument is the combination appearing in eq. \eqref{phi_def} that generates this asymmetry between matter and DM; matter can scatter into DM and not vice versa.

We should emphasize a very important point. It is clear that the Lagrangian \eqref{Lagrangian} is time-symmetric, hence the asymmetry explained above does not originate from a fundamental time-asymmetry in the theory. The reason behind this asymmetry bears similarity to the thermodynamical argument on why macroscopic processes are not time-reversal. In both cases, the time-asymmetry does not come from a fundamental time-asymmetry in nature, but from phase space considerations. The phase space of O$f$DM particles is infinitely larger compared to the matter, so a reverse process of transitioning from O$f$DM to matter is infinitely unlikely. For more discussion on this see \cite{Saravani:2015rva}.

We finish this section by mentioning two important points from phenomenological point of view.
First, the interaction involving only matter fields in the Lagrangian \eqref{Lagrangian} is $V(\phi_0)$. This, in fact, tells us that knowing the physics of matter is enough to include O$f$DM. There is no other interaction term between matter and O$f$DM in this model, which greatly restricts the model. If we know the Lagrangian of matter, we can {\it uniquely} include O$f$DM. In other words, understanding matter interactions will {\it force} O$f$DM interactions. The only new parameters are $\tilde \rho_k$ which will be discussed in the next section.

Secondly, we can generalize this idea to massive and beyond scalar fields. One can verify that all the arguments above work for massive fields as well and there is nothing particular about the matter field $\phi_0$ being massless; introduce a tower of massive O$f$DM particles on top of the mass of the matter particle,
\beq
m_k^2 = m_0^2 + k\Delta,
\eeq
where $m_0$ is the mass of $\phi_0$. It is important that the masses of O$f$DM particles are higher, otherwise the matter particles would be unstable. We must iterate that what differentiates matter from DM is how they enter the combination defined in eq. \eqref{phi_def} and not the masses.

Finally, we extend this idea to a universe with multiple fields including fermions straightforwardly. For each matter field $\psi_0$, introduce a tower of O$f$DM particles $\psi_k$ with mass $m_k^2 = m_0^2 + k\Delta$ and replace any interaction involving $\psi_0$ with $\psi = \psi_0 + \sum_k\sqrt{\tilde\rho_k \Delta}\psi_k$. This, in fact, provides a way to include O$f$DM in a more realistic matter Lagrangian like the Standard Model. In this view, $\psi_0$ is the Standard Model fields (or a subset of them). 

%%%%%%%%%%%%%%%%%%%%%%%%%%%%%%%%%%%%%%%%%%%%%%%%
%%%%%%%%%%%%%%%%%%%%%%%%%%%%%%%%%%%%%%%%%%%%%%%%
\section{Nonlocal description of O$f$DM}
%%%%%%%%%%%%%%%%%%%%%%%%%%%%%%%%%%%%%%%%%%%%%%%%
%%%%%%%%%%%%%%%%%%%%%%%%%%%%%%%%%%%%%%%%%%%%%%%%
So far, we have introduced O$f$DM and provide arguments to support that it is a viable DM candidate. However, the choices we have made in the previous section may seem arbitrary. Why there is a tower of massive particles? And why the interaction term is mediated through the particular combination in eq. \eqref{phi_def}? We answer these question in what follows and provide a physical interpretation for $\tilde \rho_k$ parameters.

Let us consider the Lagrangian \eqref{Lagrangian}. This is, in fact, a local description of a nonlocal field theory. In order to see this, consider the following path integral 
\beq\label{path_integral}
Z = \int e^{iS[\phi_0,\phi_1,\cdots]} \mathcal{D}\phi_0\prod_{k=1}^{\infty}\mathcal{D}\phi_k,
\eeq
where
\beq
S[\phi_0,\phi_1,\cdots]=\int \frac{1}{2}\phi_0 \Box \phi_0 + \sum_{k=1}^{\infty} \frac{1}{2}\phi_k(\Box-m_k^2)\phi_k - V(\phi),
\eeq
and $\phi$ is defined through eq. \eqref{phi_def}.

Irrespective of the exact form of the interaction term, we can reduce the path integral \eqref{path_integral} to an integral over only one field configuration (see \cite{Saravani:2018rwm} for detailed discussion), as follows
\beq\label{path_integral2}
Z = N \int \mathcal{D}\phi~ e^{iS_{nl}[\phi]},
\eeq
where $N$ is a numerical factor, 
\beq\label{nonlocal_action}
S_{nl}[\phi] = \int \frac{1}{2}\phi\tilde \Box_F\phi - V(\phi),
\eeq
and $\tilde \Box_F$ is defined as
\bea
\tilde \Box_F^{-1} =&& \frac{1}{\Box+i\epsilon} + \sum_{k=1}^{\infty}\frac{\tilde\rho_k\Delta}{\Box-m_k^2+i\epsilon}\notag\\
=&&\frac{1}{\Box+i\epsilon} + \int dm^2\frac{\tilde\rho(m^2)}{\Box-m^2+i\epsilon}.
\eea
Eq. \eqref{nonlocal_action} is the nonlocal action describing the same physical system \cite{Saravani:2018rwm}. In fact, as we have mentioned earlier, the nonlocal wave propagation is the starting point of studying nonlocal field theories arising from continuum approximation of wave propagation on causal sets \cite{Sorkin:2007qi, Aslanbeigi:2014zva, Saravani:2015rva}. 

In this letter, we have chosen to introduce the theory through its equivalent local description (eq. \eqref{Lagrangian}), since a local theory is more familiar to the reader and, more importantly, it provides a clearer explanation of why certain excitations of the theory behave like DM. However from a fundamental point of view, the action \eqref{nonlocal_action} is the starting point. If we start with the action \eqref{nonlocal_action} and go (in reverse) from eq. \eqref{path_integral2} to eq. \eqref{path_integral}, the origin of the tower of massive particles and the particular combination eq. \eqref{phi_def} in the previous section becomes clear.

The nonlocal description above also provides a physical explanation for $\tilde \rho(m^2)$ (or alternatively $\tilde \rho_k$). This function controls the degree of nonlocality in the action \eqref{nonlocal_action}. In fact, as $\tilde \rho(m^2)\rightarrow 0$ we recover a local massless scalar field theory. From Causal set point of view, the action \eqref{nonlocal_action} describes a local massless field theory in {\it low energies}, and the nonlocal effects are only visible when we probe such high energy scales that the inherent nonlocality of a causal set manifests itself \cite{Sorkin:2007qi}. In particular, as long as we are far below this high energy scale, $\tilde \rho(m^2)$ can be approximated by $l^2$ \cite{Saravani:2016enc} where $l$ is a length scale close to the discreteness scale of a causal set, presumed to be $l_p = \sqrt{\frac{\hbar G}{c^3}}$. From phenomenological considerations, one huge advantage of this approximation is that for all practical purposes it reduces one free function to one free variable.

%%%%%%%%%%%%%%%%%%%%%%%%%%%%%%%%%%%%%%%%%%%%
%%%%%%%%%%%%%%%%%%%%%%%%%%%%%%%%%%%%%%%%%%%%
\section{Testing O$f$DM}
%%%%%%%%%%%%%%%%%%%%%%%%%%%%%%%%%%%%%%%%%%%%
%%%%%%%%%%%%%%%%%%%%%%%%%%%%%%%%%%%%%%%%%%%%
Up until now, we have introduced O$f$DM, discussed how to include O$f$DM in matter Lagrangians and 
explained the connection to spacetime nonlocality. Here, we discuss briefly how to test this model. 

As we have discussed in the previous sections, O$f$DM particles are non-scattering, i.e. we cannot directly detect O$f$DM particles in scattering experiments. Then, according to O$f$DM model, all direct searches would fail to detect DM particles. Here, we present possible avenues that have been considered in the previous works.

\subsubsection{missing energy}
In the scattering of matter particles, there is always a chance of producing O$f$DM particles. Since O$f$DM particles are invisible to us, this would manifest itself in form of missing energy. For example if we use the approximation $\tilde\rho(m^2) = l^2$, for the Lagrangian \eqref{Lagrangian} with $V(\phi) = \lambda \phi^4$, the probability of missing energy in $2\rightarrow2$ scatterings is given by $l^2 E^2$ where $E$ is the centre of mass energy of incoming particles \cite{Saravani:2016enc}. This argument can straightforwardly be extended to more complicated Lagrangians and interactions.

\subsubsection{cosmology}
If we assume that O$f$DM constitutes a significant portion of cosmological DM, then we can use cosmological constraints to test this model. Depending on the production process of O$f$DM, this model could connect DM physics to inflation and reheating, e.g. see \cite{Saravani:2016enc}.

Moreover, cosmological O$f$DM is non-thermal since O$f$DM particles are non-scattering. This means that the (dark) matter power spectrum in this model is significantly different from thermal scenarios \cite{Saravani:2016enc}.

\subsubsection{lab experiments}
Another possibility is to test this model through laboratory experiments. See \cite{Belenchia:2016sym} for corrections to atom decay time and \cite{Saravani:2018gqu} for modifications to Casimir force induced by nonlocality.

%%%%%%%%%%%%%%%%%%%%%%%%%%%%%%%%%%%%%%%%%%%%
%%%%%%%%%%%%%%%%%%%%%%%%%%%%%%%%%%%%%%%%%%%%
\section{Summary and Conclusion}
%%%%%%%%%%%%%%%%%%%%%%%%%%%%%%%%%%%%%%%%%%%%
%%%%%%%%%%%%%%%%%%%%%%%%%%%%%%%%%%%%%%%%%%%%
In this letter, we have introduced O$f$DM model and discussed why it is a viable candidate for DM. In addition, we have explained the connection between the model and the notion of spacetime nonlocality. Finally, we have mentioned a few possible ways to test the model.

O$f$DM is fundamentally different from other DM candidates, since it does not ``postulate'' the existence of new fields to explain the DM problem. In this model, the existence of DM is a ``byproduct'' of fundamental spacetime nonlocality in nature. We have mentioned Causal Set as one possible explanation for the source of nonlocality.

We have briefly mentioned the previous works on testing O$f$DM model.
In our view, the most interesting feature of O$f$DM model is the fact that it is controlled by a single free parameter $l$.

The studies on O$f$DM is no way near complete and there are much to be done on theoretical and experimental fronts. Given the state of current direct and indirect searches for DM, we hope this letter motivates more physicists to consider O$f$DM as a viable candidate for DM and explore possible ways of testing this model.

\begin{acknowledgments}
MS is supported by the Royal Commission for the Exhibition of 1851.

\end{acknowledgments}

\bibliography{dark_matter}
\end{document}